# OPTIMIZING FINANCIAL DATA ANALYSIS: A COMPARATIVE STUDY OF PREPROCESSING TECHNIQUES FOR REGRESSION MODELING OF APPLE INC.'S NET INCOME AND STOCK PRICES

**Kevin Ungar**
Lucian Blaga University of Sibiu, Faculty of Economic Sciences, Sibiu, Romania
E-mail: kevin.ungar@ulbsibiu.ro

**Camelia Oprean-Stan**[*]
Lucian Blaga University of Sibiu, Faculty of Economic Sciences, Sibiu, Romania
E-mail: camelia.oprean@ulbsibiu.ro

**Abstract:** This article presents a comprehensive methodology for processing financial datasets of Apple Inc., encompassing quarterly income and daily stock prices, spanning from March 31, 2009, to December 31, 2023. Leveraging 60 observations for quarterly income and 3774 observations for daily stock prices, sourced from Macrotrends and Yahoo Finance respectively, the study outlines five distinct datasets crafted through varied preprocessing techniques. Through detailed explanations of aggregation, interpolation (linear, polynomial, and cubic spline) and lagged variables methods, the study elucidates the steps taken to transform raw data into analytically rich datasets. Subsequently, the article delves into regression analysis, aiming to decipher which of the five data processing methods best suits capital market analysis, by employing both linear and polynomial regression models on each preprocessed dataset and evaluating their performance using a range of metrics, including cross-validation score, MSE, MAE, RMSE, R-squared, and Adjusted R-squared. The research findings reveal that linear interpolation with polynomial regression emerges as the top-performing method, boasting the lowest validation MSE and MAE values, alongside the highest R-squared and Adjusted R-squared values.

---

[*] Corresponding author: Camelia Oprean-Stan. *E-mail: camelia.oprean@ulbsibiu.ro*

Ungar, K., Oprean-Stan, C., (2025)
*Optimizing Financial Data Analysis: A Comparative Study of Preprocessing Techniques for Regression Modeling of Apple Inc.'s Net Income and Stock Prices*

# 1. Introduction

Stock prices serve as critical indicators of a company's financial health and market sentiment. In today's dynamic financial landscape, accurately predicting stock prices is imperative for investors, analysts, and policymakers alike. Understanding the intricate relationships between a company's financial performance and its stock prices is pivotal for making well-informed investment decisions and optimizing portfolio strategies. Amidst the plethora of analytical tools available, regression analysis emerges as a potent technique for unraveling these complex relationships.

This study delves into exploring the correlation between Apple Inc.'s quarterly net income and its corresponding daily close stock prices. By scrutinizing quarterly net income data alongside daily stock prices, the aim is to find the most suitable method for data processing and forecasting large datasets from the capital market. To achieve this objective, advanced regression modeling techniques and innovative preprocessing methods were employed. Leveraging the flexibility and efficiency of Python programming, the financial datasets underwent preprocessing, utilizing techniques such as aggregation, interpolation (including linear, polynomial, and cubic spline), and lagged variables. These preprocessing methods play a pivotal role in enhancing the usability of the datasets and extracting meaningful insights for subsequent analysis.

The research makes noteworthy contributions to the field of financial analysis and regression modeling. Firstly, by conducting a comparative analysis of different preprocessing techniques, it offers valuable guidance on selecting the most suitable approach for regression modeling of financial data. Secondly, the study provides insights into the predictive power of regression models in forecasting stock prices based on quarterly net income, thereby assisting investors in making informed decisions. Additionally, the research bridges gaps in the existing literature by exploring the effectiveness of various preprocessing techniques in financial data analysis, particularly within the realm of stock price prediction.

Previous research by Hyndman and Athanasopoulos (2018) and Montgomery et al. (2015) has explored the application of time series analysis and forecasting techniques for financial data. A comprehensive body of research has analyzed the dynamics of stock markets, focusing on either the efficient market hypothesis (EMH) or the fractal market hypothesis (FMH). For instance, Oprean and Tănăsescu (2014) showed that the market efficiency hypothesis and the random walk hypothesis are no longer relevant if stock returns exhibit long-range dependence. According to Brătian et al. (2021), when the examined markets exhibit fractality, the geometric fractional Brownian motion is more appropriate for predicting stock market indexes than the geometric Brownian motion. Oprean et al. (2014) examined whether the selected capital markets adhere to a specific evolution pattern or the random walk hypothesis. These studies provide valuable insights into modeling and predicting financial trends. However, a gap exists in comprehensively evaluating the impact of various data preprocessing techniques on the effectiveness of regression models for analyzing the relationship between a company's financial health (quarterly income) and its stock prices. While Greene (2008) explores the econometric approach to efficiency analysis, it doesn't delve specifically into the selection of optimal data preprocessing methods for regression modeling in the context of stock price prediction using company income data. By addressing this gap, this research contributes to the advancement of knowledge in the field, providing practitioners with valuable guidance on selecting the most suitable preprocessing approach for regression modeling of financial data.

The focus on income rather than net profit or other variables was chosen because income reflects the total earnings generated from Apple's core business activities, which directly shows market demand and operational performance. Income has a direct impact on investor perceptions and stock prices, as it indicates the company's ability to generate sales. In contrast, net profit includes expenses, taxes, and other financial elements that might not directly correlate with stock performance, potentially introducing additional variability unrelated to the company's operational success.

The structure of the article is the following: Section I presents the literature review related to the role of data preprocessing in enhancing the performance of analytical models; sections II and III present the data collection and processing and highlight the methodology employed; section IV presents the results obtained by using the regression analysis, exploring linear and polynomial models; the last two sections discuss the findings and conclude the study.

## 2. Literature Review

The role of data preprocessing in enhancing the performance of analytical models is well-documented across various domains, including financial data analysis. This literature review explores key studies that have investigated the impact of preprocessing techniques on model performance and highlights how the current research addresses gaps and expands on these findings.

Preprocessing techniques play a crucial role in enhancing the accuracy of regression modeling in financial data analysis. These techniques address issues such as missing values, outliers, and data normalization, which can significantly impact the performance of predictive models. Recent empirical studies have demonstrated the effectiveness of various preprocessing methods in improving model accuracy, particularly in the context of financial data. Preprocessing is essential for cleaning and preparing data, which directly affects the accuracy of machine learning models.

Techniques such as data normalization, standardization, and handling missing values are fundamental to achieving reliable results (Gulati & Raheja, 2021). In financial risk analysis, preprocessing methods like mean analysis for missing values and outlier treatment have been shown to improve accuracy significantly, achieving up to 91% accuracy in some cases (Kalvala & Ahmad, 2024). Missing values can severely impact model performance. Regression-based methods for predicting missing values have been found to outperform traditional methods like mean imputation, particularly in financial datasets (Shastry et al., 2022). Outlier detection and treatment are also critical. Techniques such as max-mean analysis have been effective in improving model accuracy by addressing outliers in financial data (Kalvala & Ahmad, 2024).

Feature selection and encoding are vital preprocessing steps that influence model performance. For instance, frequency encoding has been shown to perform well with complex datasets, although its effectiveness varies with data complexity (Johnson et al., 2023). The choice of feature selection method can significantly impact the model's predictive performance, with some methods like XGBoost's importance by gaining consistent results (Johnson et al., 2023). Empirical studies have demonstrated that preprocessing can lead to substantial improvements in model accuracy. For example, preprocessing increased the accuracy of a support vector machine model from 68.7% to 88.5% (Seid & Pooja, 2019). The effectiveness of preprocessing is not limited to classification models; it is equally crucial for regression models, where it helps in achieving more accurate predictions by ensuring data quality (Gulati & Raheja, 2021). Luengo et al. (2020) elaborated on big data preprocessing techniques, offering detailed methods for dealing with voluminous datasets. Their work is particularly relevant as it provides a technical foundation for preprocessing large financial datasets using Python, as done in the current study. The techniques covered, including aggregation and interpolation, are critical for preparing the data for effective regression modeling.

While preprocessing is undeniably beneficial, it is important to note that the choice of techniques should be tailored to the specific characteristics of the dataset and the problem at hand. Not all preprocessing methods are universally applicable, and their effectiveness can vary based on the data's complexity and structure (Johnson et al., 2023). Therefore, a deep understanding of preprocessing techniques and their appropriate application is essential for optimizing model performance in financial data analysis. Table 1 summarizes some empirical studies focusing on the preprocessing techniques for data analysis and regression modeling.

Ungar, K., Oprean-Stan, C., (2025)
*Optimizing Financial Data Analysis: A Comparative Study of Preprocessing Techniques for Regression Modeling of Apple Inc.'s Net Income and Stock Prices*

**Table 1 Selection of empirical studies focusing on the preprocessing techniques for data analysis and regression modeling**

| Studies | Results | Research gap |
|---|---|---|
| Alshdaifat et al. (2021) | The performance of classification algorithms is significantly affected by the chosen preprocessing techniques. | The study identifies a gap in the automatic selection of preprocessing techniques tailored to different datasets. |
| Deari and Ulu (2023) | The TOM effect holds for most individual stocks in the MBI10 index, particularly those with higher market capitalization weights. | The study is limited to the top 10 stocks in the MSE and does not account for other factors like macroeconomic conditions or events. |
| Çetin and Yıldız (2022) | Techniques like noise filtering, missing value imputation, and feature selection are shown to significantly improve accuracy. | The study identifies a need for more research on integrating multiple preprocessing techniques and developing adaptive methods for diverse data types. |
| Famili et al. (1997) | Proper preprocessing techniques, like filtering and noise reduction, significantly enhanced data analysis in semiconductor manufacturing and aerospace applications. | The need for expert systems to guide preprocessing and the iterative nature of preprocessing, where improper techniques might lead to the loss of valuable information. |
| Heryán et al. (2024) | Higher ownership concentration generally leads to lower liquidity but does not negatively affect profitability in the long term. | Narrow focus on the automotive industry and the exclusion of other critical variables, such as innovation or external macroeconomic shocks. |
| García et al. (2016) | Preprocessing techniques like feature selection and imbalanced data handling have been successfully adapted for big data. | Scaling instance reduction methods for larger datasets and combining multiple preprocessing techniques for better performance, with a need for more work on real-time processing and non-standard learning paradigms. |
| Tosan et al. (2023) | Frequency encoding performed best for complex datasets while missing indicator imputation was the most effective for missing value handling. | Its focus is solely on XGBoost models and binary classification tasks, which limits the generalizability of the findings to other machine learning algorithms and data types. |
| Revathi and Ramyachitra (2023) | Training dataset percentage affected final classification results. | Limited discussion on the impact of dataset size variations; imbalanced dataset classification |
| Neagu and Neagu (2024) | The study found that all variables positively contributed to green development, with human capital showing the most significant effect. | The research does not explore the dynamic interactions between the determinants over shorter periods, nor does it consider the impact of real-time financial data analysis. |
| Seid and Pooja (2019) | Preprocessing significantly improved the performance of the Support Vector Machine (SVM) classifier. | The manual nature of the data collection process and the focus on a single machine learning model (SVM). |
| Gulati and Raheja (2021) | Data preprocessing significantly impacts ML algorithm generality performance. | Limited focus on consequences of data processing techniques; lack of emphasis on consequences of data processing techniques. |
| Lopata, A. et al. (2021) | The paper presents the primary results of data cube dimensions fill. | Limited data regarding company processes and analysis. |
| Barwary and Abazari (2019) | The first difference transformation was most effective at making data stationary, while the EWMA transformations performed variably across asset classes. | The chosen transformations may not generalize well across all types of financial data. |
| Kalvala and Ahmad (2024) | 91% accuracy in missing value treatment; and 92% accuracy in outlier analysis. | Data trimming leads to insufficient and inaccurate data for analysis. |
| Aditya Shastry et al. (2022) | Polynomial regression outperformed other models in predicting missing values. | Lack of comparison with other advanced regression techniques; limited discussion on potential limitations of regression-based methodology. |

Source: authors' processing.

The aim of this paper is to address specific gaps in the existing research on data preprocessing techniques, particularly within financial contexts. While previous studies have predominantly focused on the impact of preprocessing methods on classification models, there is limited exploration of their effects on regression models. This study seeks to fill this gap by systematically evaluating how different preprocessing techniques, such as interpolation and regression modeling, influence the performance of regression models in predicting stock prices based on quarterly net income. The focus is on optimizing financial data analysis and identifying the most effective preprocessing method for regression modeling.

To expand upon the current literature, this study undertakes a comparative analysis of multiple preprocessing techniques, including aggregation, linear interpolation, polynomial interpolation, cubic spline interpolation, and lagged variables. By employing these methods, the study aims to provide comprehensive insights into the techniques that most effectively enhance the predictive power of regression models in the context of financial data. This comparative analysis will contribute to a more robust understanding of the preprocessing methods that optimize the performance of predictive models, specifically in the financial domain.

The research also explores the correlation between Apple Inc.'s quarterly net income and its daily closing stock prices. By analyzing these two datasets together, the study aims to uncover the most suitable data processing and forecasting method for large financial datasets, with a particular emphasis on capital market analysis.

Based on this analysis, the following hypothesis is proposed:

$H_0$: *Among the various preprocessing techniques applied to Apple Inc.'s financial datasets, linear interpolation followed by polynomial regression provides superior predictive performance in modeling Apple Inc.'s net income and stock prices, compared to other preprocessing methods.*

This hypothesis will guide the comparative evaluation of preprocessing techniques, with the goal of identifying the optimal approach for financial data forecasting.

## 3. Methodology
### 3.1 Data Collection and Processing

This section outlines the methodology employed to process the financial datasets of Apple Inc., including quarterly income and daily close stock prices, into five distinct datasets. The original time frames for the datasets were as follows: the close stock prices (3774 observations) were available from January 2, 2009, to December 29, 2023, while the quarterly income data (60 observations) covered the period from March 31, 2009, to December 31, 2023. The evolution in time of the two variables can be observed in the figures below:

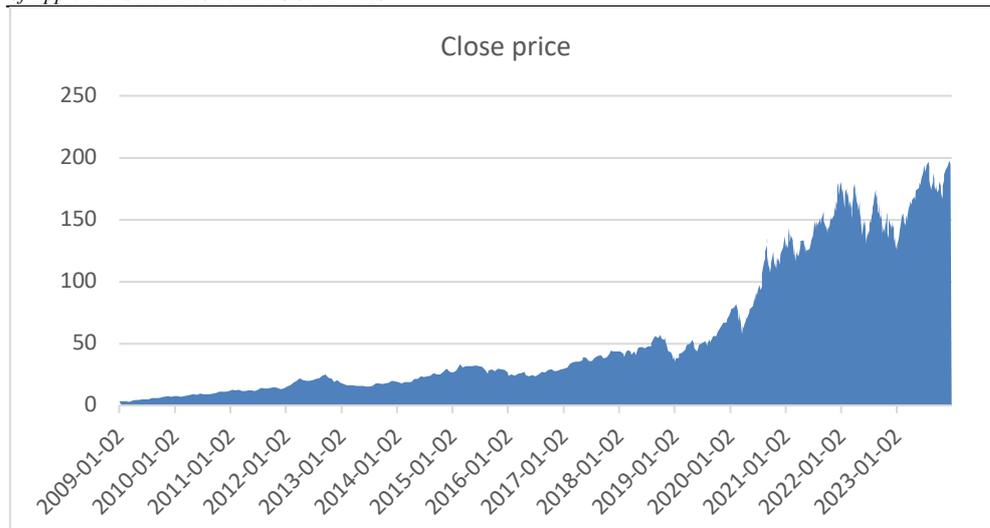

**Figure 1 Apple close stock prices**
Source: authors' processing.

Figure 1 shows a steady rise in Apple's stock price from 2009 to 2019, followed by a sharp surge with notable fluctuations from 2019 onward. The significant increase starting in 2019 coincides with the onset of the COVID-19 pandemic, which accelerated demand for technology products and services.

While Apple's stock price exhibited a steady rise from 2009 to 2019 followed by a sharp surge from 2019 onward (Figure 1), Figure 2 illustrates that Apple's quarterly income has shown more consistent growth with periodic fluctuations. The income increased gradually from 2009 to 2019 but surged significantly after the pandemic began, reaching a peak of $34.6 billion in Q4 2021. This post-2019 rise in income aligns with the stock's upward trend, driven by heightened demand for Apple's products and services during the pandemic.

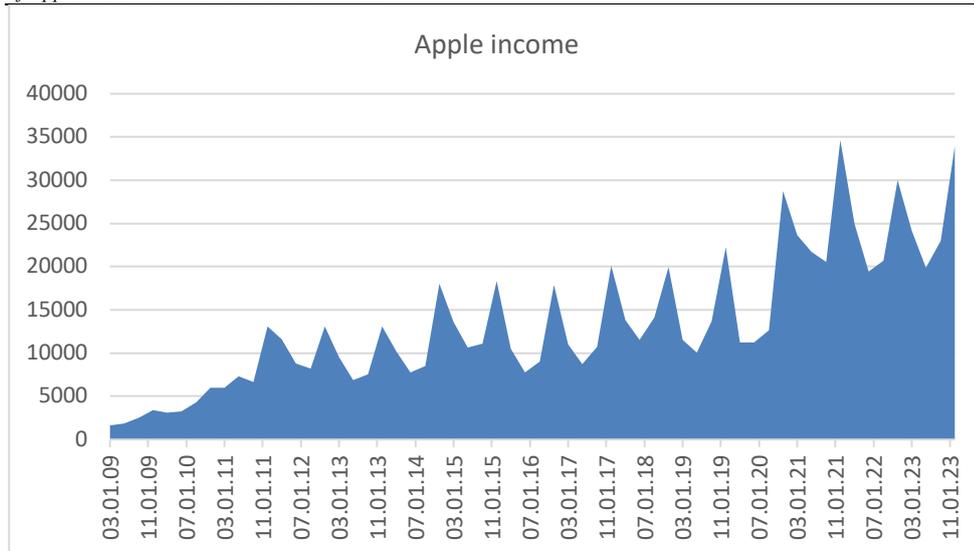

**Figure 2 Apple's quarterly net income**
Source: authors' processing.

Quarterly results are often reported after the end of the quarter. In this analysis, the reporting dates for financial data, as listed on Yahoo Finance and MacroTrends, were used to align the timing of the data with the corresponding impact on stock prices. The data collection for the independent variable "Apple Quarterly Net Income" is sourced from Macrotrends, while Apple Inc. (AAPL) close stock prices are obtained from Yahoo Finance. Although stooq.com provides quarterly quotes, the study aimed to align daily stock prices with quarterly financial results. This approach allowed daily fluctuations to be modeled and the relationship with quarterly revenue data to be better understood.

Apple's quarterly income is measured in million US dollars with a total of 60 observations and 3774 observations for daily stock prices (measured in US dollars). After preprocessing and merging, the final time frame for the datasets spans from March 31, 2009, to December 31, 2023, because during the preprocessing phase, rows where the independent variable "Apple Quarterly Net Income" had values but the dependent variable "Apple Inc. (AAPL) close stock prices" did not (or vice versa) were dropped to ensure alignment between the two variables.

Apple Inc. (AAPL) was chosen due to its significant stock market impact and consistent financial reporting, which provides a solid data set for analysis. In addition, Apple is a widely followed and analyzed company, which makes its stock behavior a topic of interest for many investors and researchers.

Based on Apple's history during the reviewed period (2009/03/31 to 2023/12/31), no major corporate operations, such as mergers or acquisitions, were identified that would have significantly impacted Apple's core operations in a way that would alter the analysis of stock prices and quarterly earnings trends during this time frame.

These datasets are crafted using different preprocessing techniques aimed at enhancing usability and extracting valuable insights for subsequent analysis. Through detailed explanations of each method: aggregation, interpolation (linear, polynomial, and cubic spline) and lagged variables, the article will provide a comprehensive overview of the steps taken to transform raw data into analytically rich datasets.

### 3.1.1 Aggregation

When dealing with large datasets with varying frequencies and timeframes (Hyndman & Athanasopoulos, 2018), such as quarterly income data and daily stock prices, it's often necessary to aggregate the data to make it more manageable and suitable for analysis. Aggregation is the process of combining multiple data points into a smaller set of summary statistics, typically over a larger timeframe or at a lower frequency. The process of the research was as follows:

The first step in aggregation was aligning the time intervals of both datasets. Since the income data is reported quarterly and the stock prices are reported daily, the daily stock prices were aggregated into quarterly intervals to match the income data.

To aggregate the daily stock prices into quarterly intervals, the average (mean) daily stock price for each quarter was calculated. For example, to calculate the average stock price for the first quarter (Q1), the sum of all the daily stock prices for that quarter was divided by the number of trading days in that quarter.

Once the stock prices are aggregated into quarterly intervals, the data are merged with the quarterly income data. This merging process aligns the corresponding income and stock price data for each quarter, creating a unified dataset suitable for analysis.

### 3.1.2 Linear interpolation

Linear interpolation (Montgomery et al., 2015) is a method used to estimate values between two known data points, assuming a linear relationship between the data points and calculating intermediate values accordingly. The process involves determining the equation of the line connecting two adjacent data points and then extrapolating the desired value based on this linear relationship. Mathematically, in this context, the two known data points are (x1, y1) and (x2, y2), where x represents time (date) and y denotes the corresponding value (income or stock price). In the research, this method was applied to both Apple's quarterly net income (independent variable) and Apple's daily close stock price (dependent variable).

The reason for applying linear interpolation to both variables was the need to align the daily stock prices with the quarterly income data and to ensure a more comprehensive dataset for analysis. The original dataset for Apple's daily stock price (AAPL), sourced from Yahoo Finance, had 3774, while Apple's quarterly income had only 60 observations. After applying the linear interpolation, the number of observations increased to 5387. Linear interpolation filled missing data points based on trends between surrounding values, ensuring data continuity. This method was applied not only for trading days but also for any missing data points within the range, regardless of whether they occurred on weekdays or holidays. This approach allowed for a more complete dataset and enabled a better analysis of trends in both Apple's quarterly net income and daily stock prices.

When the linear interpolation was applied to the two data frames using Python the numbers of observations per variable were slightly different, but by merging the two sets, the total amount of observations was 5387. This approach of having the same number of observations per variable was applied for both polynomial and cubic interpolation and the final number of observations was the same as the linear interpolation in the end.

The linear interpolation formula applied using Python can be written as:

$$x = x1 + \left(\frac{x2 - x1}{y2 - y1}\right) x (y - y1) \tag{1}$$

Where:
− x1 and x2 represent the known quarterly income values or known daily price values. These are essentially the income values available at times $y1$ and $y2$. The formula used for both independent and dependent variables is the same, so based on the context, but in the following explanation will be used the quarterly values for the explanation of the process applied.
− $y1$ and $y2$ represent the time points (dates) corresponding to the income values $x1$ and x2.

The formula aims to estimate the income value $x$ at a given time $y$ lying between $y1$ and y2 by assuming a linear relationship between time and income. It calculates the difference between the given time $y$ and the first time point $y1$, denoted as $y-y1$. This difference represents the proportion of time elapsed between $y1$ and $y$. The formula scales the time difference $y-y1$ by the ratio of the differences between income values and time points: $(x2-x1)/(y2-y1)$. This ratio represents the rate of change in income per unit change in time. Multiplying it by the time difference yields an estimate of the change in income corresponding to the elapsed time. The scaled difference is added to the initial income value $x1$ to obtain the estimated income value $x$ at the given time $y$.

In Python, the interpolation process was achieved by using the "interpolation=linear" argument in functions such as "numpy.interp()". So, the mathematical formula described above outlines the underlying principles of the linear interpolation used in the programming code.

### 3.1.3 Polynomial interpolation

Similar to linear interpolation, polynomial interpolation estimates values between known data points by assuming a polynomial relationship. It calculates the coefficients of a polynomial function based on the given data points, allowing for the approximation of values at intermediate points within the range (Press, 2007). In this research, a second-degree polynomial function was employed to fit the given data points.

Second-degree polynomial interpolation, also known as quadratic interpolation, utilizes a polynomial of the form:

$$f(x) = ax^2 + bx + c \qquad (2)$$

To interpolate the quarterly income data, the objective is to calculate the coefficients $a$, $b$, and $c$ that minimize the error between the interpolated values and the actual income data. For each known data point $(xi, yi)$, an equation based on the polynomial function above is constructed, resulting in a system of equations. This system can be represented in matrix form as:

$$\begin{bmatrix} x1^2 & x1 & 1 \\ x2^2 & x2 & 1 \\ . & . & . \\ . & . & . \\ Xn^2 & Xn & 1 \end{bmatrix} \begin{bmatrix} a \\ b \\ c \end{bmatrix} = \begin{bmatrix} y1 \\ y2 \\ . \\ . \\ Yn \end{bmatrix} \qquad (3)$$

Solving this system of equations using methods like Gaussian elimination or matrix inversion yields the coefficients $a$, $b$, and $c$ of the quadratic polynomial. Once these coefficients are determined, the polynomial function can be used to estimate intermediate values for any given time between the known data points. But in Python, instead of those 2 methods, were employed numerical algorithms tailored for efficiency and stability, encapsulated within high-level functions provided by the library NumPy, as numpy.polyfit() for polynomial interpolation.

Polynomial interpolation of degree 2 provides a more flexible curve than linear interpolation, allowing for capturing more complex patterns in the data.

### 3.1.4 Cubic spline interpolation

Similar to the first two interpolation methods mentioned earlier, cubic spline interpolation (Press, 2007) is a technique used to estimate values between known data points by fitting cubic polynomials to segments between these points. It ensures smoothness and continuity by imposing conditions on the derivatives of adjacent polynomials at the interval boundaries. The process involves dividing the dataset into intervals and fitting a cubic polynomial to each interval. The resulting curve is continuous, with matching first and second derivatives at interval boundaries, ensuring smoothness and providing a reliable framework for estimating values between data points, facilitating accurate analysis and decision-making. The whole process for realizing the cubic interpolation can be divided into 5 steps:

*Dividing the Dataset*

Given a set of data points $(x_i, y_i)$, where $x_i$ represents the independent variable (time) and $y_i$ represents the dependent variable (Apple net income), the dataset is divided into intervals $[x_i, x_i+1]$.

*Fitting Cubic Polynomials*

Within each interval, a cubic polynomial of the form below is fitted to the data points. Here, $a_i$, $b_i$, $c_i$, and $d_i$ are the coefficients of the polynomial:

$$P_i(x) = a_i + b_i(x - x_i) + c_i(x - x_i)^2 + d_i(x - x_i)^3 \qquad (4)$$

*Ensuring Continuity*

To ensure continuity and smoothness, the cubic spline interpolation imposes conditions on the derivatives of adjacent polynomials at the interval boundaries. Specifically:

- **Continuity of Function**: The interpolated function $P_i(x)$ is continuous at each boundary point $x_i$. This means that the value of the curve at $x_i$ on the left side of the point should be the same as the value on the right side. This condition ensures that the value of the interpolated function $P_i(x)$ at the boundary point $x_i$ is the same from both sides of the boundary.
- **Continuity of First Derivative**: The first derivative of the interpolated function $P_i'(x)$ is continuous at each boundary point $x_i$. This ensures that the slope of the curve at $x_i$ from both sides should be the same. Continuity of the first derivative means the steepness of the curve should be consistent where two segments meet, preventing abrupt changes in the curve's steepness.
- **Continuity of Second Derivative**: The second derivative of the interpolated function $P_i''(x)$ is continuous at each boundary point $x_i$. This ensures consistent curvature across segment boundaries, resulting in a smooth and visually pleasing curve. It ensures that the change in the steepness of the curve is smooth where two segments meet.

These conditions together ensure that the interpolated curve looks smooth and natural at the boundaries of each segment, creating a system of equations that can be solved to determine the coefficients of the cubic polynomials. By enforcing continuity of function, first derivative, and second derivative, cubic spline interpolation creates a seamless connection between adjacent segments, resulting in a visually pleasing and mathematically robust interpolation method.

*Interpolation*

Once the coefficients are determined, the interpolated values for any given *x* within an interval can be calculated using the corresponding polynomial $Pi(x)$.

In Python, to perform cubic spline interpolation on a data frame, the following line of code was used: "df_income.resample('D').interpolate(method='cubic')".

This interpolates the values in the data frame to the daily frequency, encapsulating the essence of the entire explained process within that line of code.

### 3.1.5 Lagged variables

This method involves shifting a time series dataset backward by a certain number of time periods (Hyndman & Athanasopoulos, 2018). In simpler terms, it means looking at the past values of a variable to predict its future behavior. This technique is particularly useful in examining the dynamic relationship between two variables over time. In the research, the dataset was shifted backward by 1 and at the end the two datasets were merged together, resulting in 42 observations per variable. The 'Apple income' dataset had 60 observations and the 'close price' data frame had 3774 observations. After creating the lagged versions of the variables by shifting the values down by one row, the "NaN" values are introduced for the first row of each lagged variable because there's no previous value to lag. During the merge operation based on the 'date' column, rows with "NaN" values are dropped from the lagged dataset. Additionally, if there are dates present in one dataset but not in the other, or if there are missing values for the lagged variables on certain dates, those rows will also be dropped during the merge operation. So, the 42 observations in the final dataset are a result of missing values in the original datasets.

### 3.2 Regression Analysis: Exploring Linear and Polynomial Models

This section will uncover insights about the regression models, delving into the metrics that verify which model performs better through a comparison of the performance. The evaluation criteria will include the following metrics: cross-validation score MSE (Validation MSE), MSE, MAE, RMSE, R-squared, and Adjusted R-squared (Greene, 2008).

*Validation MSE (cross-validation score)* - the cross-validation score is computed as the average of the negative MSE values across all folds. Negative MSE is calculated for each fold of the cross-validation process. This approach facilitates consistent interpretation of model performance, where less negative scores indicate superior accuracy. For example, a cross-validation score of -0.21 is considered better than -0.31, reflecting a more accurate model. The choice of using negative MSE aligns with scoring conventions where higher scores indicate better performance. This nuanced approach ensures effective assessment and optimization of machine learning models in real-world applications.

*MSE (Mean Squared Error)* - MSE is calculated by taking the average of the squared differences between predicted values and actual values. It measures the discrepancy between estimated and actual values. While validation scores offer a broad evaluation, the MSE test provides a detailed analysis of how well the model performs on unseen data.

*MAE (Mean Absolute Error)* - is calculated by taking the average of the absolute differences between predicted values and actual values. It measures the average magnitude of errors in predictions, providing a robust assessment of model performance. The MAE test provides general insights on prediction accuracy, but compared to MSE (Mean Squared Error), MAE is less sensitive to outliers (it doesn't amplify the influence of data points that fall far away from the main cluster of data through squaring). While MSE penalizes larger errors more heavily due to squaring, MAE treats all errors equally. Squaring the equation amplifies the differences between small and large errors, so large errors have a more significant impact on the final MSE result. This makes MAE particularly useful when the dataset contains significant outliers or when the absolute magnitude of errors is more important than their squared values.

*RMSE (Root Mean Squared Error)* - is calculated by taking the square root of the average of the squared differences between predicted values and actual values. Compared to MAE and MSE, taking the square provides a measure that is in the same unit as the original target variable. This can make interpretation more intuitive as it represents the typical size of the errors in the same scale as the target variable. Additionally, RMSE penalizes larger errors more heavily than MAE, making it sensitive to outliers similar to MSE. However, unlike MSE, RMSE has the advantage of being interpretable in the original units of the target variable, which can aid in understanding the practical implications of the model's performance.

In all 3 cases (MSE, MAE, RMSE), the lower the score is (close to 0), the better the model performs.

*R-squared* - measures the proportion of the variance in the dependent variable that is predictable from the independent variables in a regression model. Variance represents how much the values in a dataset differ from the mean value. It ranges from 0 to 1, where a value of 1 indicates that the model explains all the variability of the response data around its mean. In other words, the closer the R-squared is to 1, the better the model fits the data. Higher R-squared values indicate a better fit of the regression model to the data, suggesting that the model can effectively explain and predict the observed outcomes. However, it's important to note that R-squared alone does not provide information about the correctness of the model's assumptions, the presence of multicollinearity, or the potential for overfitting. Therefore, R-squared should be interpreted in conjunction with other diagnostic measures and considerations specific to the analysis at hand.

*Adjusted R-squared* - while R-squared provides a measure of how well the model fits the data, R-squared adjusted takes into account the number of predictors in the model, providing a more accurate assessment of the model's goodness of fit. Predictors are the independent variables or features used in the regression model to predict the dependent variable. R-squared adjusted penalizes the inclusion of unnecessary variables in the model, thus addressing the issue of overfitting. Unlike R-squared, which can artificially increase with the addition of more predictors, R-squared adjusted accounts for the degrees of freedom and only increases if the addition of a predictor improves the model fit beyond what would be expected by chance. Therefore, R-squared adjusted is a more reliable metric for assessing model performance, especially when comparing models with different numbers of predictors. Additionally, R-squared adjusted is useful even when the model has only one independent variable because it provides a more conservative estimate of the model's goodness of fit. This helps prevent overestimation of the model's explanatory power and ensures that the reported fit is not artificially inflated by chance, ultimately helping to prevent overfitting.

### 3.2.1 Linear regression model in python

The building of the linear regression model (Scikit-learn, n.d.) started with the extraction of the independent variable x from the 'Apple income' column of the data frame, and reshaped to a 2D array. Then the dependent variable y is extracted from the 'Close price' column. After the extraction of x and y, each variable was split into 2 sets, X_train and y_train which represent 75% of each dataset and X_test and y_test which represent 25% of the total dataset. Both X and y were shuffled before splitting, to prevent biases or pre-built trends in the training and testing sets. This ensures a more representative sample for training the model. Both X and y were scaled using the z score normalization method ("method = StandardScaler ()"). After the scaling of the data, X_train and y_train were fitted in a linear regression model ("model = LinearRegression()"). The purpose of the split into 2 datasets is to train the model with 75% of the data frame and test the result using the 25% parts, which represent the unseen data for the model, to test its generalization capacity. The trained model is used to predict the target variable (y_prediction) on the scaled testing data X_test. For the linear regression model X_test represents an unseen input of data in the model. Once y_prediction was computed, all the test scores explained earlier in the article were calculated using mostly: y_test, y_prediction and X_test (for the Adjusted R-squared). For the visual representation of the result, a scatter plot ("plt.scatter()") was used with X_test and y_test as arguments and for the line ("plt.plot()") which represents the regression model, X_test and y_prediction were used.

### 3.2.2 Polynomial regression model in python

The steps for the creation of the polynomial model were similar to the linear regression model, with the exception of the feature (predictor) transformation process (the independent variable X). After the splitting of each dataset into train and test sets, the independent variables X_train and X_test were transformed into "X_train_poly" and "X_test_poly", respectively, incorporating polynomial features. The original data (ex: X_train) had one feature (represented by a single value in each inner list). After applying the transformation (poly.fit_transform()), the transformed data X_train_poly now has three features. The first feature remains the same (the original values), and the two additional features represent the polynomial terms (square of the original features in this case). This method is useful for capturing non-linear relationships between features in the dataset that might not be evident with the original features alone.

Only the independent variable obtained polynomial features because the aim was to capture potential non-linear relationships between the independent variable and the target variable (y). The transformation allows the model to consider higher-order interactions between the features, enabling it to better fit the data and potentially improve predictive performance compared to a linear model. Meanwhile, the dependent variable (y) remains unchanged as it does not undergo any transformation in the polynomial regression process.

### 4. Results of Regression Analysis and Model Evaluation

The results of regression analysis and model evaluation are given in Table 2. The interpretation of these results is given below.

Table 2 Results of regression analysis and model evaluation

| Data processing technique | Regression models | Validation MSE | MSE | MAE | RMSE | R-squared | Adjusted R-squared |
|---|---|---|---|---|---|---|---|
| Aggregation | Linear | -0.29911 | 0.51155 | **0.60215** | **0.71523** | 0.60658 | 0.57632 |
| | Polynomial | -0.30845 | **0.54552** | 0.58070 | 0.73859 | 0.58045 | 0.46603 |
| **Linear interpolation** | Linear | -0.26548 | 0.24517 | 0.37253 | 0.49514 | 0.74899 | 0.74880 |
| | **Polynomial** | **-0.24911** | 0.22801 | **0.33665** | 0.47750 | **0.76656** | **0.76603** |
| Polynomial interpolation | Linear | -0.31603 | 0.29263 | 0.40082 | 0.54095 | 0.70056 | 0.70033 |
| | Polynomial | -0.30680 | 0.28371 | 0.37862 | 0.53264 | 0.70969 | 0.70904 |
| Cubic spline interpolation | Linear | -0.31825 | 0.29468 | 0.40240 | 0.54284 | 0.69841 | 0.69819 |
| | Polynomial | -0.30872 | 0.28562 | 0.38033 | 0.53443 | 0.70768 | 0.70703 |
| **Lagged variables** | Linear | -0.33470 | 0.22445 | 0.38992 | 0.47376 | **0.42670** | 0.36300 |
| | Polynomial | **-0.35071** | 0.22197 | 0.38467 | 0.47114 | 0.43302 | **0.19003** |

Source: authors' processing.

*Aggregation*:
- **Linear regression**: The model has a validation MSE of -0.29911, indicating that, on average, the squared difference between predicted and actual values during cross-validation is approximately 0.29911 units. The MSE, MAE, and RMSE values for the test set are 0.51155, 0.60215, and 0.71523, respectively, which are relatively high. The R-squared and Adjusted R-squared values of 0.60658 and 0.57632 indicate that approximately 60.66% of the variance in the dependent variable (stock prices) is explained by the independent variable (quarterly income).
- **Polynomial regression**: The model exhibits a slightly higher validation MSE of -0.30845 compared to linear regression. The MSE, MAE, and RMSE values for the test set are 0.54552, 0.58070, and 0.73859, respectively. The R-squared and Adjusted R-squared values are 0.58045 and 0.46603, suggesting a slightly weaker fit compared to linear regression. Polynomial regression may capture non-linear relationships better, but in this case, it doesn't seem to offer significant improvement over linear regression.

*Linear interpolation:*
- **Linear regression**: The model demonstrates a lower validation MSE of -0.26548 compared to aggregation. The MSE, MAE, and RMSE values for the test set are 0.24517, 0.37253, and 0.49514, respectively, indicating better performance. The R-squared and Adjusted R-squared values of 0.74899 and 0.74880 suggest a strong linear relationship between quarterly income and stock prices, with approximately 74.90% of the variance explained.
- **Polynomial regression**: The model exhibits a validation MSE of -0.24911, slightly lower than linear regression. The MSE, MAE, and RMSE values for the test set are 0.22801, 0.33665, and 0.47750, respectively, showing slightly better performance than linear regression. The R-squared and Adjusted R-squared values of 0.76656 and 0.76603 indicate a slightly better fit compared to linear regression.

*Polynomial interpolation*:
- **Linear regression:** The model has a validation MSE of -0.31603, slightly higher than linear interpolation. The MSE, MAE, and RMSE values for the test set are 0.29263, 0.40082, and 0.54095, respectively. The R-squared and Adjusted R-squared values of 0.70056 and 0.70033 suggest a good fit, but slightly lower than linear interpolation.
- **Polynomial regression:** The model exhibits a validation MSE of -0.30680, similar to linear regression. The MSE, MAE, and RMSE values for the test set are 0.28371, 0.37862, and 0.53264, respectively. The R-squared and Adjusted R-squared values of 0.70969 and 0.70904 indicate a slightly better fit compared to linear regression.

*Cubic spline interpolation*:
- **Linear regression**: The model demonstrates a validation MSE of -0.31825, slightly higher than polynomial interpolation. The MSE, MAE, and RMSE values for the test set are 0.29468, 0.40240, and 0.54284, respectively. The R-squared and Adjusted R-squared values of 0.69841 and 0.69819 suggest a good fit, but slightly lower than polynomial interpolation.
- **Polynomial regression**: The model exhibits a validation MSE of -0.30872, similar to the linear regression. The MSE, MAE, and RMSE values for the test set are 0.28562, 0.38033, and 0.53443, respectively. The R-squared and Adjusted R-squared values of 0.70768 and 0.70703 indicate a slightly better fit compared to linear regression.

*Lagged variables:*
- **Linear regression**: The model has the highest validation MSE of -0.33470 among all techniques, indicating poorer performance. The MSE, MAE, and RMSE values for the test set are 0.22445, 0.38992, and 0.47376, respectively. The R-squared and Adjusted R-squared values of 0.42670 and 0.36300 suggest a weaker fit compared to other techniques.
- **Polynomial regression**: The model exhibits a validation MSE of -0.35071, slightly higher than linear regression. The MSE, MAE, and RMSE values for the test set are 0.22197, 0.38467, and 0.47114, respectively. The R-squared and Adjusted R-squared values of 0.43302 and 0.19003 indicate the poorest fit among all techniques.

Based on the results, linear interpolation with polynomial regression performs the best, as it has the lowest validation MSE and MAE values, along with the highest R-squared and Adjusted R-squared values. This suggests that the polynomial regression model with linear interpolation provides the most accurate predictions of stock prices based on quarterly income. To better represent the result, figure 3 shows the graph of the polynomial model.

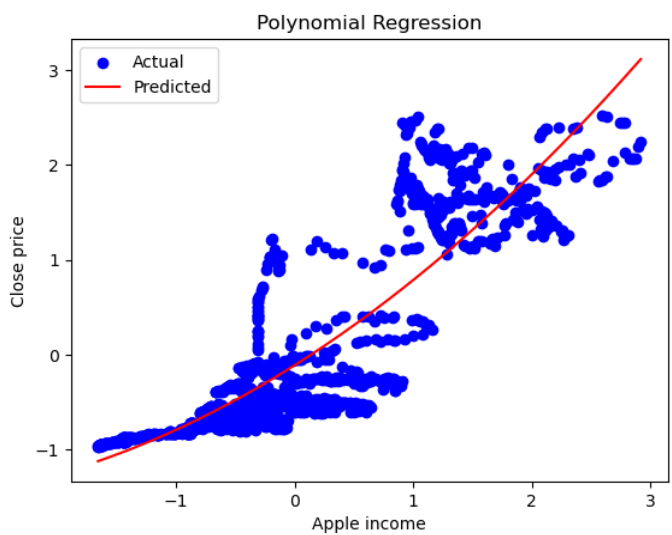

**Figure 3 Polynomial regression (degree 2) using the linear interpolation technique**
Source: authors' processing.

The values depicted in the graphs were standardized using the standardization method. This ensures that the analysis is not unduly influenced by the disparate unit scales of the variables. For instance, the independent variable (Apple income) was measured in US million dollars, while the dependent variable (Close price) was measured in US dollars.

Conversely, lagged variables with polynomial regression perform the worst, with the highest validation MSE, as well as the lowest Adjusted R-squared values. This indicates that the polynomial regression model with lagged variables poorly explains the variance in stock prices based on quarterly income. Despite lagged variables with polynomial regression showing the lowest MSE and RMSE, it's still considered the worst-performing model due to its lower Adjusted R-squared value. To better represent the differences between the two results, figure 4 is shown a graphical representation.

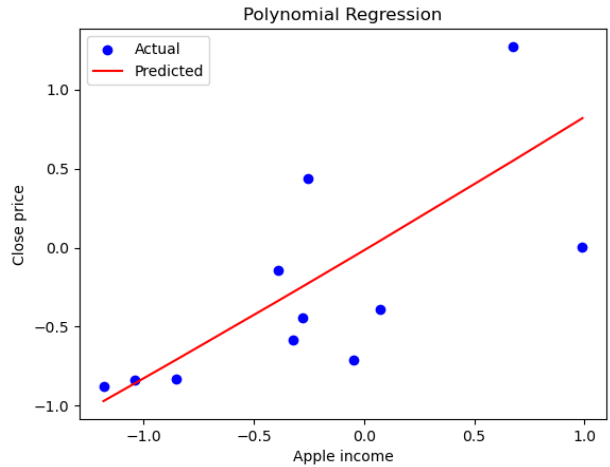

**Figure 4 Polynomial regression using the lagged variables technique**
Source: authors' processing.

These performance disparities stem from the effectiveness of each data processing technique in capturing the underlying relationship between quarterly income and stock prices, as well as the respective number of observations per technique. The graphs presented above exhibit fewer observations than the total number available per method (42 for lagged variables and 5387 for linear interpolation). This discrepancy arises because the graphs were constructed using the scaled X_test and y_test portions of the dataset, representing only 25% of the total observations. The remaining 75% of observations from both datasets were utilized to train the polynomial model (X_train and y_train).

In both graphs, the blue points (Actual) represent only 25% of the observations, while the red line (Predicted) depicts the polynomial regression model fitted on the remaining 75% of the observations per variable.

The aim of the model is to forecast the dependent variable, although it yields unsatisfactory results in both cases, despite the polynomial regression using the linear interpolation technique performing better than others. Now that it was identified the superior technique and model, we seek to enhance the performance of the polynomial regression model. This involves gradually modifying the equation, increasing its complexity from the second degree until it begins to overfit. The resulting modifications are detailed in Table 3.

**Table 3 Performance comparison of polynomial regression models with increasing complexity**

| Model | Degree | Validation MSE | MSE | MAE | RMSE | R-squared | Adjusted R-squared |
|---|---|---|---|---|---|---|---|
| polynomial regression | 2 | -0.24911 | 0.22801 | 0.33665 | 0.47750 | 0.76656 | 0.76603 |
| | 3 | -0.22614 | 0.20253 | 0.30840 | 0.45003 | 0.79264 | 0.79202 |
| | 4 | -0.21996 | 0.19636 | 0.29818 | 0.44313 | 0.79896 | 0.79821 |
| | 5 | -0.21319 | 0.18866 | 0.30192 | 0.43435 | 0.80684 | 0.80598 |
| | 6 | -0.19900 | 0.17303 | 0.29001 | 0.41597 | 0.82284 | 0.82192 |
| | 7 | -0.19610 | 0.17127 | 0.29732 | 0.41385 | 0.82465 | 0.82360 |
| | 8 | -0.17518 | 0.15139 | 0.26209 | 0.38909 | 0.84500 | 0.84396 |
| | 9 | -0.17358 | 0.15003 | 0.26768 | 0.38734 | 0.84639 | 0.84524 |
| | 10 | -0.16358 | 0.14138 | 0.24886 | 0.37601 | 0.85524 | 0.85405 |
| | 15 | -0.15668 | 0.13445 | 0.24016 | 0.36668 | 0.86234 | 0.86068 |
| | ***32*** | ***-0.21099*** | ***0.12665*** | ***0.23176*** | ***0.35589*** | ***0.87032*** | ***0.86706*** |
| | 33 | -0.20733 | 0.12832 | 0.23445 | 0.35822 | 0.86862 | 0.86521 |
| | 34 | -0.96958 | 0.93629 | 0.80363 | 0.96762 | 0.04142 | 0.01583 |

Source: authors' processing.

Overall, the adjusted R-squared value being smaller than 0.9 indicates that while the model captures a significant portion of the variance in the data, it is not a perfect fit. This suggests that other factors beyond Apple's quarterly income may influence stock prices. This does not imply a lack of correlation but rather that the model, as constructed, includes some noise or unexplained variability.

It's evident from the table that as the degree of the polynomial increases, the model's complexity grows, leading to a gradual improvement in the R-squared and adjusted R-squared values up to degree 32. However, beyond this point, a decline in both metrics is observed, indicating the onset of overfitting. Notably, the model's performance shows significant improvement from degree 2 to 10, but from degree 11 onwards, the rate of improvement begins to slow down. To streamline the presentation and better highlight the evolution of the polynomial model, the calculation of values between the range of degrees 10 to 32 was skipped. This decision helps avoid overcomplicating the table while still capturing the essential trends in model performance. To provide a visual comparison of the model changes, a graph depicting polynomial regression of degree 32 is presented in Figure 5.

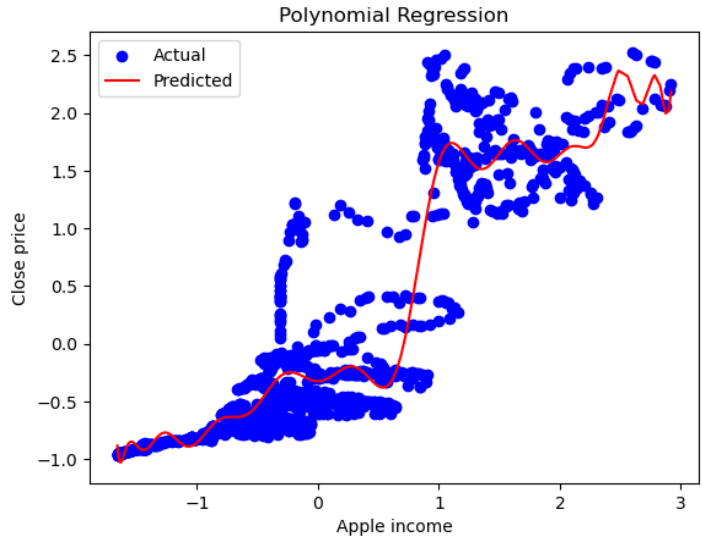

**Figure 5 Polynomial regression (degree 32) using the linear interpolation technique**
Source: authors' processing.

Figure 5 demonstrates a notable improvement compared to the one with degree 2. As the degree increases, the polynomial model attempts to closely fit all data points on the graph, thereby enhancing its ability to forecast the outcome. The reason why increasing the degree of the polynomial model provided a better fit to the data is because higher-degree polynomials have more flexibility to capture complex patterns and variations in the data. As the degree increases, the model becomes more adept at capturing intricate relationships between the independent and dependent variables, allowing it to better approximate the underlying structure of the data. This increased flexibility enables the model to minimize the errors between the observed data points and the predicted values, resulting in improved predictive performance and a closer alignment between the model's predictions and the true values. Mathematically, the red line (Predicted) in the graph above can be represented as the following equation:

$$y = b0 + b1x + b2x^2 + b3x^3 + \cdots + b32x^{32} \qquad (5)$$

Where:
- y is the predicted value of the dependent variable ("Close price"),
- x is the independent variable ("Apple income"),
- b0, b1, b2, b3......b32 are the coefficients of the polynomial regression model.

Each coefficient is estimated independently during the model fitting process. So, after the coefficient of $x$ is computed, it doesn't mean that all subsequent coefficients will be the same. Each coefficient is unique to its corresponding term in the polynomial equation and will have a different value.

The equation of the polynomial regression model represents the predicted relationship between the independent variable ("Apple income") and the dependent variable ("Close price"). This equation defines the predicted line that the model aims to fit to the observed data points, which comprise a subset of the dataset (X_test and y_test) used for generalization. When plotted on a graph, the data points represent the observed values, while the overlaying predicted line generated by the polynomial regression model represents the best-fitting approximation of the underlying relationship between the variables, as per the model's perspective. The coefficients in the equation determine the shape and orientation of this line. In regression analysis, the objective is to estimate these coefficients in a manner that minimizes the discrepancy between the predicted values and the actual observed values, thus achieving the most accurate representation of the relationship between the variables. A polynomial of degree 32 for modeling the relationship between income and stock prices was chosen primarily for demonstration purposes rather than practical application. The high degree showcases the flexibility and capacity of polynomial regression to fit complex, non-linear patterns in data. However, such a model is more theoretical and serves to illustrate how well the model can adapt to the data points, even at the risk of overfitting. In practice, simpler models are often preferred for their interpretability and generalizability to unseen data. The use of a high-degree polynomial highlights the balance needed between fitting data and maintaining a model that is both understandable and useful in real-world applications.

## 5. Discussions

The study's findings highlight the critical role of data preprocessing techniques in financial data modeling, particularly in the context of regression analysis. The results confirm that linear interpolation followed by polynomial regression offers superior performance in predicting Apple Inc.'s stock prices based on quarterly net income. This outcome aligns with prior research that underscores the importance of interpolation methods in enhancing model performance, as found in studies focusing on financial time series forecasting. However, while past studies have emphasized the importance of preprocessing for classification models, this study extends these insights to the domain of regression modeling in financial contexts, filling a notable gap in the literature.

The high predictive power demonstrated by linear interpolation in combination with polynomial regression, particularly in terms of its R-squared and Adjusted R-squared

STUDIA UNIVERSITATIS ECONOMICS SERIES
"Vasile Goldiș" Western University of Arad

Ungar, K., Oprean-Stan, C., (2025)
*Optimizing Financial Data Analysis: A Comparative Study of Preprocessing Techniques for Regression Modeling of Apple Inc.'s Net Income and Stock Prices*

values, confirms that simpler interpolation methods can outperform more complex ones. In contrast, other methods, such as the lagged variables and aggregation methods, showed significantly lower performance, particularly in terms of explanatory power, indicating limitations in capturing the relationship. The lagged variables method, with its limited observation count, struggled to capture the relationship between net income and stock prices. Similarly, the aggregation method, despite having more observations, exhibited lower R-squared values than the interpolation techniques. Interestingly, the aggregation method with both the linear and polynomial regression models also demonstrated suboptimal performance, as indicated by their lower R-squared and Adjusted R-squared values compared to other techniques. Notably, the lagged variables method had 42 observations per variable in its dataset, while the aggregation method had 60 observations per variable. In contrast, all three interpolation methods had a total of 5387 observations per variable. These findings underscore the importance of dataset size in capturing the true relationship between independent and dependent variables. The discrepancy in performance between methods with varying observation counts highlights the significance of sample size in regression analysis. These findings emphasize the importance of dataset size in regression analysis and the superiority of interpolation methods in leveraging larger datasets to capture the relationship between variables more effectively.

The study's results regarding the suboptimal performance of the lagged variables method and aggregation methods are consistent with previous research indicating that smaller datasets or improper aggregation may dilute the predictive power of regression models (in line with Pourkamali-Anaraki & Hariri-Ardebili, 2023; O'Neill & Costello, 2023; Darik et al., 2024; Guo et al., 2022; Gründler & Krieger, 2022). In financial markets, where high-frequency data are common, the importance of larger datasets is critical for capturing the intricate relationships between independent and dependent variables (Pardy et al., 2018). The fact that the interpolation methods performed best with larger datasets further supports this conclusion, in line with Lux et al. (2021) and Lanz et al. (2022).

Moreover, the inferior performance of the cubic spline and polynomial interpolation methods compared to linear interpolation suggests potential overcomplexity in these techniques. While cubic spline and polynomial interpolation methods may offer flexibility, their additional complexity may not necessarily translate into improved model performance. In many cases, simpler models, such as linear interpolation, may suffice for capturing the underlying relationships in the data. Thus, careful consideration of model complexity is essential in regression analysis to ensure optimal performance and interpretability.

The polynomial regression model of degree 32 demonstrated a notable improvement in performance metrics, with both R-squared and Adjusted R-squared values increasing by approximately 11%, from 76% to 87%. While this increase may initially suggest a stronger relationship between Apple's quarterly net income and daily stock prices, it's crucial to delve deeper into the implications of such a high-degree polynomial model. Despite the apparent improvement in explanatory power, the complexity introduced by a polynomial regression of degree 32 raises concerns about overfitting. Overfitting occurs when a model fits the training data too closely, capturing noise or random fluctuations rather than the true underlying relationships between variables. In our context, this would imply that the polynomial regression model may be overly sensitive to fluctuations in Apple's quarterly net income, leading to inflated performance metrics on the training data but poor generalization to new, unseen data. This finding contrasts with studies that argue for the greater flexibility of more sophisticated interpolation techniques (for example, Jiang et al., 2024 or Raubitzek & Neubauer, 2021), suggesting that model complexity must be carefully managed to avoid overfitting.

Investigating alternative predictors or supplementary datasets may enhance the model's predictive power and reduce the risk of overfitting. By addressing these aspects, future research can offer deeper insights into the effectiveness and limitations of polynomial regression models for forecasting stock prices in the context of financial analysis.

## 6. Conclusions and Recommendations

This study aimed to evaluate the effectiveness of various preprocessing techniques applied to Apple Inc.'s financial datasets for regression modeling of net income and stock prices. The findings confirmed the hypothesis that linear interpolation followed by polynomial regression provides superior predictive performance compared to other preprocessing methods. This conclusion is supported by the fact that the linear interpolation with polynomial regression achieved the lowest validation Mean Squared Error (MSE) and Mean Absolute Error (MAE), indicating higher accuracy in predicting stock prices based on quarterly income data. Additionally, this technique exhibited the highest R-squared and Adjusted R-squared values, suggesting strong explanatory power and minimal risk of overfitting the data. The study also revealed that while cubic spline and polynomial interpolation methods may offer increased flexibility, their added complexity did not result in better model performance compared to the simpler linear interpolation technique. This suggests that simpler models, like linear interpolation, may be more suitable for financial data analysis by striking a balance between model complexity and interpretability. The concerns raised about the performance of the polynomial regression model of degree 32 reinforce this argument, as excessive model complexity led to concerns about overfitting and sensitivity to random fluctuations. Linear interpolation followed by polynomial regression outperformed other approaches such as cubic spline interpolation, lagged variables, and aggregation, confirming the hypothesis and reinforcing the importance of choosing the right preprocessing methods in financial data analysis. Simpler models, such as linear interpolation, strike a balance between performance and complexity, providing strong predictive accuracy without overfitting.

The scientific novelty of the article lies in the application of polynomial regression to model the relationship between income and share price over a certain period and in the comparison of different interpolation methods (linear, polynomial, cubic) to fill the missing data gaps. While Matlab and other software programs offer more adaptation functions, this study uniquely combines financial data analysis with interpolation techniques to improve dataset completeness and predictive modeling. The originality of this research lies in the systematic comparison of various interpolation methods combined with polynomial regression for financial data modeling. This study provides valuable insights for improving data preprocessing and predictive modeling, particularly in financial contexts.

Given the findings, this study recommends further exploration of simpler preprocessing methods in financial data modeling. Researchers and practitioners should prioritize techniques like linear interpolation when dealing with large datasets, as they provide a combination of accuracy and interpretability. Additionally, caution should be exercised when applying high-degree polynomial regression models, as these may lead to overfitting, especially in noisy financial datasets. Future studies should also consider incorporating additional predictors, such as external macroeconomic factors or supplementary financial data, to enhance the robustness of regression models. Exploring more advanced machine learning models and their interaction with different preprocessing methods could provide new insights into optimizing predictive performance in financial contexts.

**Acknowledgments**
The authors thank the anonymous reviewers and editor for their valuable contribution.

**Funding**
This research received no funding.

STUDIA UNIVERSITATIS ECONOMICS SERIES
"Vasile Goldiş" Western University of Arad

Ungar, K., Oprean-Stan, C., (2025)
*Optimizing Financial Data Analysis: A Comparative Study of Preprocessing Techniques for Regression Modeling of Apple Inc.'s Net Income and Stock Prices*

**Authors Contribution**

Conceptualization, K.U. and C.O.S.; methodology, K.U.; software, K.U.; validation, K.U. and C.O.S.; formal analysis, C.O.S.; investigation, K.U.; data curation, K.U.; writing—original draft preparation, K.U.; writing—review and editing, C.O.S; supervision, C.O.S.; project administration, C.O.S.

**Disclosure Statement**

The authors have not any competing financial, professional, or personal interests from other parties.